\documentclass[pop,fleqn]{w-art}
\usepackage{times}
\usepackage{amsfonts}

\def\cL{{\cal L}}
\def\bR {\mathbb{R}}

\begin{document}

\pagespan{1}{}



\title[CTC removal and supertubes]{Removal of closed time-like curves 
by supertube domain walls}

\author[N. Drukker]{Nadav Drukker\footnote{
     e-mail: {\sf drukker@nbi.dk}}}
     \address{The Niels Bohr Institute, Copenhagen University\\
     Blegdamsvej 17, DK-2100 Copenhagen, Denmark.}

\begin{abstract}
  We discuss how closed time-like curves can be eliminated from 
  certain supergravity backgrounds by inclusion of domain-walls 
  made of supertubes. Special emphasis is given to the mechanism 
  by which the supertubes spread into domain walls, which is 
  similar to the enhan\c{c}on mechanism.\\
  Lecture notes from my talk at the RTN meeting in Kolymbari, Crete,
  September 2004.
\end{abstract}
\maketitle                   

\section{Introduction}

Closed time-like curves, or ``time machines'' pose an intriguing 
puzzle in theoretical physics. While generally considered to be 
unphysical (including by this author), they are common in solutions 
of general relativity. The standard practice in the field has usually 
been to simply discard such solutions.

In recent years quite a few solutions of supergravity were found 
that resemble G\"odel's universe \cite{Goedel} and are highly 
supersymmetric \cite{Gauntlett, Harmark}. 
This led to a small industry of finding more such solutions, and there 
were some proposals on how string theory may deal with spaces 
suffering from this pathology \cite{Horava}.

Given that string theory incorporates general relativity with 
the rest of physics, including quantum mechanics, it seems like the 
correct laboratory to address the issue of time machines. So the 
question we address here is 
whether those spaces are correct solutions of string theory, or 
is there a mechanism in string theory that changes those metrics in 
a way that removes the pathology.

The main claim is that, in the few examples studied, those supergravity 
solutions are {\em not} legitimate solutions of string theory. This 
is demonstrated by considering certain D-brane objects (supertubes 
\cite{Mateos}) in these background. Those supertubes may be 
taken as the building blocks of those geometries, and their 
dynamics cease to make sense when their compact direction wraps 
a closed time-like curve. In a close analogy with the enhan\c{c}on 
mechanism \cite{enhancon} we conclude that the supertubes go into 
a new phase before hitting the closed time-like curves. In that phase 
they delocalize and form a domain wall. Across the domain wall the 
metric is non-differentiable, so the original metrics, which are 
analytic, and their time machines are just the naive continuation 
of the metric ignoring the domain walls.

These notes are a summary of a talk given at the RTN meeting in 
Kolymbari, Crete, September 2004. More details and a full list 
of references can be found in the original papers 
\cite{we,Drukker:2004zm}.

\section{Metrics}

Our starting point is a family of metrics that solve the equations of 
motion of type IIA supergravity \cite{emparan}
\begin{equation}
  \begin{aligned}
 \label{supertube}
    ds^2=&-U^{-1}V^{-1/2}(dt-A)^2+U^{-1}V^{1/2}dy^2
          +V^{1/2}\sum_{i=1}^8 (dx^i)^2 \\
    B_2=&U^{-1}(dt-A)\wedge dy-dt\wedge dy\,, \\
    C_1=&-V^{-1}(dt-A)+dt\,, \\
    C_3=&-U^{-1}dt\wedge dy\wedge A\,, \\
    e^\Phi=&U^{-1/2}V^{3/4}\,.
  \end{aligned}
\end{equation}
Here $U$ and $V$ are harmonic functions in the eight $x^i$ 
directions and $A$ is a one-form satisfying the Maxwell equation. 
From the harmonic functions we can see that these metrics describe 
a collection of fundamental strings stretched in the $y$ direction as 
well as D0-branes. For appropriate values of the one-form, $A$, the 
metric will carry angular momentum and the three-form potential in 
the Ramond-Ramond sector indicates the existence of a D2-brane. 
Those are exactly the ingredients that make up supertubes 
\cite{Mateos,emparan}, and indeed this is the supergravity 
description of those objects.

Let us consider two examples of metrics of this class. Of the eight 
directions labeled by $x^i$ above we use radial coordinates in the 
subspace spanned by $x^1,\cdots, x^4$, and Cartesian coordinates 
in the other four. The radial coordinate is $r$ and the tree Euler 
angles on $S^3$ are $\theta$, $\psi$ and $\phi$. Throughout we 
will use the left one-forms
\begin{equation}
\begin{aligned}
&\sigma^L_1=\sin\phi\,d\theta-\cos\phi\sin\theta\,d\psi\,,
\\
&\sigma^L_2=\cos\phi\,d\theta+\sin\phi\sin\theta\,d\psi\,,
\\
&\sigma^L_3=d\phi+\cos\theta\,d\psi\,,
\end{aligned}
\end{equation}
in terms of which the metric on $S^3$ is
\begin{equation}
ds^2_{S^3}=\frac{1}{4}\left(
(\sigma^L_1)^2+(\sigma^L_2)^2+(\sigma^L_3)^2\right)\,.
\end{equation}

The first solution is given by
\begin{equation}
U=C_1\,,\qquad
V=C_2\,,\qquad 
A=\frac{1}{2}C_3\,r^2\sigma^L_3\,,
\end{equation}
Where $C_1$, $C_2$ and $C_3$ are constants.

This metric was first described in \cite{Gauntlett}, and is one of the large 
class of G\"odel like solutions of supergravity. 
While the general metric of the class (\ref{supertube}) preserves 
eight supersymmetries, this specific ansatz preserves 20. To study the 
causal structure of this space it's instructive to look at the angular 
components of the metric
\begin{equation}
\frac{\sqrt{V}}{4}r^2\left[
(\sigma_L^1)^2+(\sigma_L^2)^2
+\left(1-\frac{C_3^2}{UV}r^2\right)(\sigma^L_3)^2\right]\,.
\end{equation}
Thus for radii $r>UV/C_3^2$ there are closed time-like curves, 
given by the integral curves of the vector field $\partial_\phi$ 
dual to $\sigma^L_3$, or the Hopf fibres.

The second example is given by
\begin{equation}
U=1+\frac{Q_s}{r^2}\,,\qquad
V=1+\frac{Q_0}{r^2}\,,\qquad 
A=\frac{J}{2r^2}\sigma^L_3\,,
\end{equation}
Where $Q_s$, $Q_0$ and $J$ are constants. These harmonic functions 
indicate a source localized at the origin of this $\bR^4$ and smeared 
in the transverse $\bR^4$. This metric is the type IIA dual of the 
two charge black hole of type IIB, where the charges are usually of 
D1 and D5-branes. Here the three constants correspond to the 
fundamental string charge, the D0-brane charge density and angular 
momentum respectively.

Again we want to look at the angular part of the metric
\begin{equation}
\frac{\sqrt{V}}{4}r^2\left[
(\sigma_L^1)^2+(\sigma_L^2)^2
+\left(1-\frac{J^2}{UVr^6}\right)(\sigma^L_3)^2\right]\,.
\end{equation}
So now there will be closed time-like curves for very small radii, 
$r^6<J^2/(UV)$. For large charges this simplifies to 
$r^2<J^2/(Q_0Q_s)$.

Of those two examples the second one is asymptotically flat and the 
problem of closed time-like curves arises only near the origin. In the 
case of G\"odel there was no problem near the origin, but at large 
radii the space does not make sense. This suggests the possibility of 
patching the two metrics; replacing the singular center of the black 
hole with a piece of the G\"odel space.

This can be done for any radius $r=R$ and the metric will be continuous 
provided we take the constants
\begin{equation}
C_1=1+\frac{Q_s}{R^2}\,,\qquad
C_2=1+\frac{Q_0}{R^2}\,,\qquad 
C_3=\frac{J}{R^4}\,.
\end{equation}

\section{The domain wall}

We chose the constants describing the interior geometry such that the 
metric will be continuous, but it is not differentiable at the interface, 
indicating the existence of a source at $r=R$. It is easy to see what 
the source is, either by using the Israel 
matching conditions \cite{israel}, 
or just by Gauss's law. The G\"odel space 
solves the vacuum equations, while the black hole carries charges. 
Therefore the charges associated with the black hole are now 
located at the domain wall.

So at $r=R$ we have a domain wall that has $Q_0$ D0-brane 
density, $Q_s$ fundamental strings as well as $J$ units of angular 
momentum. For the D0-branes and fundamental strings to carry 
angular momentum they have to form a bound state, a supertube 
\cite{Mateos}. The simplest description of this composite object 
is as a D2-brane extended in the $t$ and $y$ directions and 
wrapping the Hopf fibres of the $S^3$. To get the domain wall we 
have to smear the source along all the transverse directions: the 
$S^2$ base of the fiber and the transverse $\bR^4$.

We have not specified yet the radius of the domain wall, $R$, and 
for certain choices the metric will still include closed time-like curves. 
In fact since the source is made up of supertubes there are a few 
restrictions. First the radius of the supertube is related to the charges it 
carries by $R^2=Q_0Q_s$. In addition there is a bound on the angular 
momentum $J\leq Q_0Q_s=R^2$. This is enough to guarantee that there 
are no closed time-like curves.

The analysis so far indicates that if we assume that the source of the 
black-hole is not singular and localized at the origin it will be made 
of supertubes who will have a radius $R^2\geq J$ that will prevent 
the appearance of closed time-like curves. But is there a preferred 
radius?

To answer that question we look at this background from the point of 
view of the supertubes themselves. i.e we probe the metric with another 
supertube. As mentioned, a supertube is best described as a D2-brane, 
and the fact that it carries string charge is manifested 
by an electric field $E=1$ and  the D0-brane charge density is 
proportional to the magnetic field $B$.

At low energies the dynamics of the supertube will be best 
described by a non-commutative field theory with open string metric 
$G$ and non-commutativity parameters $\Theta$ calculated from 
the closed string metric $g$, the pullback of the Neveu-Schwarz 2-form 
$B_2$ and the world-volume field strength $F$ and given by 
\cite{Kruczenski:2002mn}
(the lines and columns correspond to the $t$, $\phi$ 
and $y$ directions)
\begin{equation}
 G^{ab}+\Theta^{ab}=
 \left(\frac{1}{g+B_2+2\pi\alpha'F}\right)^{ab}=
\frac{1}{B}
\begin{pmatrix}
-m&&-V^{1/2}&&f+Vr^2/B\cr
-V^{1/2}&&0&&1\cr
-(f+Vr^2/B)&&-1&&V^{1/2}r^2/B
\end{pmatrix}\,,
\label{open-string-metric}
\end{equation}
where
\begin{equation}
m=\frac{Vg_{\phi\phi}}{B}
+\frac{V^{1/2}U}{B}\left(B+\frac{f}{U}\right)^2\,,
\qquad\hbox{and}\qquad
A=f\sigma^L_3\,.
\label{m}
\end{equation}

Slow rigid motion at velocity $v$ of the supertube in the transverse 
directions is governed by an action with kinetic term
\begin{equation}
\cL=\frac{1}{2}M\,v^2\,,
\label{action}
\end{equation}
where $M=\int d\phi\,m$ is the metric on moduli space. This 
quantity is positive if the supertube wraps a space-like curve, but 
for a certain choice of $B$ will be negative when wrapping a closed 
time-like curve, or zero when wrapping a closed null curve.

This behavior is very similar to the enhan\c{c}on mechanism 
\cite{enhancon}, where 
the metric on the moduli space of a D6-brane wrapping a $K3$ surface 
vanishes at a finite radius. There it was interpreted as a sign that the 
brane cannot continue further in to smaller radii, and instead goes into 
a new phase, where a $U(1)$ gauge symmetry is enhanced to $SU(2)$.

We wish to apply the same reasoning to the case at hand. The parameter 
$M$ corresponds to the inertial mass of the supertube for motion in the 
transverse directions. At the radius where $M=0$ there could be big 
fluctuations in those directions, and the description of 
the supertubes as localized objects breaks down. The gauge theory 
describing them should enter new phase and the supertubes are naturally 
smeared into the domain wall.

This calculation of the moduli space metric is based on the supergravity 
solution but we can trust it due to the high degree of supersymmetry, 
the supertubes preserve eight supercharges in this background. It 
would be very interesting to understand the new phase of the 
non-commutative field theory in which the supertubes are delocalized, 
but until we have this description we can trust the results from 
supergravity.

\section{Discussion}

The preceding discussion summarizes the main claim in 
\cite{Drukker:2004zm}, that supertube naturally form domain 
walls to shield against the creation of closed time-like curves in those 
backgrounds. Another example studied there is the case of the BMPV 
black hole \cite{BMPV}, or its type IIA dual, which carries three 
charges, of D0 and D4-branes and fundamental strings.

The same kind of domain wall geometry can be constructed in 
that example, where from the outside it looks like the BMPV black 
hole. But inside (behind the Cauchy horizon) the metric is replaced 
by a certain G\"odel-like space. This construction was also found 
independently by Gimon and Ho\v{r}ava \cite{Gimon:2004if}.
The necessary source for making up the domain wall in this case would 
be a bound state of the three objects whose charges are carried by 
the black hole. Such a three-charge generalization of the supertube 
was indeed found by Bena and Kraus \cite{Bena:2004wt}.

Since those objects preserve only four supercharges the moduli space 
may be corrected by quantum effects. So an argument similar to the one 
above would not be rigorous. Still we find it likely that the general 
picture will still hold, and those objects will experience an 
enhan\c{c}on like phenomenon and form the appropriate domain wall.

In the example studied here both spaces, the G\"odel-like metric and the 
two-charge black hole, had closed time-like curves. But we argued that 
neither space is a complete solution of string theory, and the correct 
solution includes a domain wall that patches those two metrics together. 
Thus each of these spaces could be considered the wrong analytical 
continuation of a legitimate solution, ignoring the domain wall.

Such a mechanism is a possible solution to many other metrics that 
have closed time-like curves. Locally those solutions are usually 
perfectly good, and problems arise only when extending the solutions 
to global metrics. Such continuations are unique assuming the metric 
is analytic, which was {\em not} the case here. Domain walls are 
a natural way to break analyticity and thus averting this global 
problem.

Another beautiful example where a similar mechanism is at play 
was found by Israel \cite{Israel:2003cx}. In that case a G\"odel-like 
deformation of $AdS_3$ develops closed time-like curves, but those 
are naturally cured after the inclusion of a domain wall made of 
large fundamental strings. It would be very interesting to see 
how general this mechanism is.

\end{document}